\documentclass[a4paper]{paper}  
\usepackage[margin=2.5cm]{geometry}
\usepackage{lipsum}
\usepackage{xcolor}
\sectionfont{\large\sf\bfseries\color{black!70!blue}} 
\usepackage{graphicx} 
\usepackage{authblk}

\usepackage{braket}
\usepackage{physics}
\usepackage{multirow}
\usepackage[algo2e]{algorithm2e} 
\usepackage{siunitx}

\usepackage{subcaption}
\usepackage{algorithm}
\usepackage{hyperref}
\usepackage{cleveref}
\usepackage{biblatex}
\bibliography{quantum-ai} 

\begin{document}
\title{\centering An inductive bias from quantum mechanics: learning order effects with non-commuting measurements}

\author[1, 2]{Kaitlin Gili}
\affil[1]{\small University of Oxford, Oxford, United Kingdom OX1 2JD }
\affil[2]{\small Xanadu, Toronto, ON, M5G 2C8, Canada}
\author[2]{Guillermo Alonso}
\author[2]{Maria Schuld}

\maketitle

There are two major approaches to building good machine learning algorithms: feeding lots of data into large models, or picking a model class with an ``inductive bias'' that suits the structure of the data. When taking the second approach as a starting point to design quantum algorithms for machine learning, it is important to understand how mathematical structures in quantum mechanics can lead to useful inductive biases in quantum models. In this work, we bring a collection of theoretical evidence from the Quantum Cognition literature to the field of Quantum Machine Learning to investigate how non-commutativity of quantum observables can help to learn data with ``order effects'', such as the changes in human answering patterns when swapping the order of questions in a survey. We design a multi-task learning setting in which a generative quantum model consisting of sequential learnable measurements can be adapted to a given task -- or question order -- by changing the order of observables, and we provide artificial datasets inspired by human psychology to carry out our investigation. Our first experimental simulations show that in some cases the quantum model learns more non-commutativity as the amount of order effect present in the data is increased, and that the quantum model can learn to generate better samples for unseen question orders when trained on others - both signs that the model architecture suits the task.

\section{INTRODUCTION}\label{s:intro}

In the quest to understand if and how quantum algorithms for machine learning (ML) will be useful, it is valuable to conduct research from more than one angle \cite{PerdomoOrtiz2017, opp_challenge_QML, Schuld_2022}. 
Typical approaches in the past years have investigated learning settings where the computational complexity of the overall algorithm  \cite{hinsche2021learnability, hinsche2022, 2020Du, Liu_2021_speed, schreiber2022classical, cai2022sample}, generalization scores on benchmarks \cite{gili2023generalization, hibatallah2023framework, gili_qcbm, moussa2023application}, or trainability guarantees \cite{Mcclear2018Barren, holmes2022expressivity, wang2021noiseinduced, cerezo2021costfunction, arrasmith2021gorges, rudolph2023trainability} were the primary focus. While these approaches have indeed pushed our knowledge boundary forward, an alternative angle is to explore the \textit{inductive bias}, or natural structure that models based on quantum mechanics exhibit \cite{ragone2023representation, Pesah_2021, bowles2023contextuality}, and find applications for quantum machine learning that suit this bias. After interesting model classes are found, questions related to classical simulatability and generalization power can be reassessed from a new vantage point. We add to this work here by constructing and studying a quantum model with a bias that results from non-commutativity in quantum measurement. 

More specifically, we introduce a learning setting inspired by the literature on ``quantum cognition'' \cite{quantum_cognition, busemeyer2012quantum, busemeyer2011quantum, question_order_quantum, Widdows_2023} which allows us to investigate whether the non-commutative nature of quantum observables encourages certain quantum models to learn data distributions that contain order effects. Order effects can be thought of as sequences of random variables whose joint distribution changes when the sequence is permuted; an example that we will consider here is the change of Yes/No answer patterns to survey questions when humans are presented with questions in different orders. To control the structure in the data (such as the strength and kind of order effect), we create artificial probability distributions of human answering patterns inspired by the human cognition and psychology literature \cite{measure_new_ordereffects, sequence_effects, wellbeing_order_effects}. 

The quantum model we consider to learn distributions over answering patterns is a generative model in which we interpret the results of a trainable two-outcome measurement as the answer to a question. An answer pattern is then a collection of $N$ Yes/No answers generated by measuring $N$ observables in sequence. The model gets adapted to a given question order by permuting the observables accordingly. 
Mathematically and conceptually, this investigation design differs fundamentally from existing generative models such as quantum Born machines \cite{Benedetti_2019, Liu_2018, Cheng2017, Gili2022} and quantum Boltzmann machines \cite{kieferova2016tomography, korenkevych2016benchmarking, Cheng2017}. Firstly, instead of measuring the state of a set of qubits, we take subsequent measurements of the same state, which can be understood as ``collapsing'' the wavefunction several times. Since such a workflow exceeds the typical blueprint of quantum algorithms, one of our contributions is the implementation of the model on a mixed-state simulator without access to sequential measurements, for which we use PennyLane \cite{bergholm2022pennylane}. Secondly, the problem we consider is a multi-task learning problem \cite{multi-task_review, multi_task_generative}, where the goal is to generalize to unseen training data \textit{as well as} to related unseen tasks. \cite{bowles2023contextuality} independently found that such a framework suits as a bridge between quantum foundations and machine learning.

We use this study design to walk the first steps of exploring two basic questions: 
\begin{enumerate}
    \item As we increase the strength of the order effect in the dataset, does the model learn a stronger degree of non-commutativity? 
    \item Can the quantum model generalize better on unseen orders after training on other orders? 

\end{enumerate}
An affirmation of both -- and we will define the criteria in more detail below -- would give support to the hypothesis that non-commutativity can create an inductive bias to learn order effects. Our results presented here indicate that on some of the datasets, the non-commutativity of the trained model grows with the strength of the order effect in the data, and that for up to five observables, training on a sufficiently large number of tasks lowers the generalization error on unseen tasks, even if we never trained on those. We view these results as an initial investigation into whether the inductive bias of non-commutativity is the primary resource driving the model to learn the underlying pattern of order effects. In our outlook, we conclude with some future research questions prompted from this study that could strengthen or weaken this claim in the future. 

We note that the study design does not aim at making any strong statements towards the performance of the quantum model compared to classical machine learning. We also do not test the approach with regards to current industry applications. This may be surprising considering the dominant culture in the quantum machine learning literature, but we believe that investigations with a limited but well-defined scope that study the qualitative behaviour of a model provide important additional insights. This will be further discussed in Section \ref{s:outlook}.

\section{THE QUANTUM LEARNING SETTING}\label{learn_setting}

In the first part of this work we introduce the learning setting, including the model and problem to solve. We provide a theoretical intuition in Section \ref{ind_bias} by discussing the nature of non-commutativity in quantum theory and its relation to structures that exist in binary distributions containing order effects - i.e. measured changes in the distribution due to the ordering of the variables. Here, we also define a measure for the \textit{amount of non-commutativity}, which we utilize to quantify the non-commutativity within the model at each training step. Building off of these insights, in Section \ref{qcircuit} we introduce a multi-task quantum generative architecture that can be used to investigate whether the structure of non-commutativity biases the model towards learning binary order effects, and provide an overview of the model's implementation. Lastly, we introduce the psychology-inspired datasets intentionally designed to carry out the study.

\paragraph{Motivating the Model Design}
\label{ind_bias}

A critical motivation of this investigation is taken from a result that is increasingly recognized in the classical machine learning literature: models that contain an \textit{inductive bias} that matches the structure in the dataset it aims to learn, will ultimately have better generalization performance with typically fewer samples \cite{goyal2022inductive, HAUSSLER1988177, feinman2018learning}. Informally, an inductive bias can be defined as the preference of a model for one hypothesis that is consistent with the training data over another. Inductive biases encourage the model towards this hypothesis, including extra information about the problem in the model itself, rather than in the training data. Examples of inductive biases in classical machine learning are priors in Bayesian Networks, translation-invariance in Convolutional Neural Networks (CNNs) or regularizers in linear regression \cite{goyal2022inductive, cohen2017inductive, oshea2015introduction}. Turning to the inductive biases of quantum models, we can leverage non-classical mathematical structures and behaviors that exist in quantum mechanics, and search for alignments within correlations in datasets \cite{bowles2023contextuality}. Here, we discuss non-commuting observable measurements in quantum mechanics, and provide theoretical intuition for its structural connection to binary distributions with order effects. This choice is inspired by a sub-field known as ``Quantum Cognition'' which utilizes elements from quantum theory to build models that explain human cognitive behavior \cite{quantum_cognition, busemeyer2012quantum, busemeyer2011quantum, question_order_quantum}. The basic hypothesis of quantum cognition is that certain features of distributions generated by human decision-making are difficult to model by classical probability theory while naturally found in the statistics of quantum mechanics \cite{question_order_quantum}. An example \cite{pothos2022quantum} is the \textit{conjunction fallacy}, or the consistent observation that humans estimate the probability of two conditions to be true -- such as the probability that a Scandinavian person has blond hair \textit{and} blue eyes -- higher than the probability of one of them to be true (i.e., the probability that a Scandinavian person has blond hair). This violates the set theoretic foundations of classical probability theory which devise $p(A \And B) \leq p(A)$. Order effects are another example that has been studied using small-scale datasets of medical and legal decision making \cite{trueblood2011quantum}. The quantum cognition literature concludes that quantum mechanics could be an interesting mathematical framework to model such effects. In some sense our study can be seen as an attempt to probe this observation from a machine learning angle. 

\subsection{Non-commutativity}

Let us review the basics of sequential measurements in quantum theory in order to motivate the design of our quantum model, a section that readers with a background in quantum theory can easily skip. Consider a set of $N$ quantum observables $\{Q_{n}\}, n=1,\dots,N$ describing measurements that can have two possible outcomes. Note that there are $N!$ permutations of these measurements which we call ``(measurement) orders''. One can view each observable $Q_{n}$ to be a question regarding binary information that we would like to know about the quantum system. Without loss of generality we assume that each observable $Q_{n}$ of arbitrary size only contains the eigenvalues $\{-1, 1\}$ and denote by $\{|u^{n}_{1}\rangle, ..., |u^{n}_{m}\rangle\}$ the eigenstates associated with $+1$ and by $\{|v^{n}_{1}\rangle, ..., |v^{n}_{l}\rangle\}$ the eigenstates associated with $-1$. Let $\ket{\psi_0}$ be a quantum state that we wish to measure with respect to the observable, such that we can express it using the eigenbasis of $Q_n$: 

\begin{equation}
    |\psi_{0}\rangle =  \sum_{i}\alpha_{i}|u^{n}_{i}\rangle + \sum_{i}\beta_{i}|v^{n}_{i}\rangle
\end{equation}\label{initial_state}

After the measurement the state $| \psi_1 \rangle$ of the system is given by $\frac{1}{\sqrt{\sum_{i}|\alpha_{i}|^2}}\sum_{i}\alpha_{i}|u^{n}_{i}\rangle$ with probability $\sum_{i}|\alpha_{i}|^2$ and by $\frac{1}{\sqrt{\sum_{i}|\beta_{i}|^2}}\sum_{i}\beta_{i}|v^{n}_{i}\rangle$ with probability $\sum_{i}|\beta_{i}|^2$. 

Now consider a second observable $Q_m$, and express the eigenstates of $Q_n$ in this new eigenbasis, $ |u^{n}_{i}\rangle =  \sum_{j}\gamma^{u}_{ij}|u^{m}_{j}\rangle + \sum_{j}\zeta^{u}_{ij}|v^{m}_{j}\rangle $, and 
$|v^{n}_{i}\rangle =  \sum_{j}\gamma^{v}_{ij}|u^{m}_{j}\rangle + \sum_{j}\zeta^{v}_{ij}|v^{m}_{j}\rangle$. This allows us to rewrite $|\psi_1 \rangle$ in the eigenbasis of $Q_m$, and -- after cancelling some terms -- obtain the following probability distribution over the two measurement outcomes:

\begin{align}\label{p_nm}
P_{nm}(+1, +1) &= \sum_{j} \sum_{i}\left|\alpha_{i}\gamma_{ij}^{u}\right|^2 \\
P_{nm}(-1, +1) &= \sum_{j}\sum_{i}\left|\beta_{i}\gamma_{ij}^{v}\right|^2 \\
P_{nm}(+1, -1) &= \sum_{j} \sum_{i}\left|\alpha_{i}\zeta_{ij}^{u}\right|^2\\
P_{nm}(-1, -1) &= \sum_{j}\sum_{i}\left|\beta_{i}\zeta_{ij}^{v}\right|^2.
\end{align}

If the two Hermitian observables $Q_{n}, Q_{m}$ share common eigenspaces, they can be written as $Q_{n} = VD_{n}V^{\dagger}$ and $Q_{m} = VD_{m}V^{\dagger}$, where $V$ is a unitary and $D_{\{n,m\}}$ a diagonal operator. It then follows that

\begin{gather}
    Q_{n} Q_{m} = VD_{n}V^{\dagger}VD_{m}V^{\dagger} = 
    VD_{n}D_{m}V^{\dagger} = \\ VD_{m}D_{n}V^{\dagger}  = 
    \nonumber VD_{m}V^{\dagger}VD_{n}V^{\dagger} = Q_{m}Q_{n}
\end{gather}

When the observables do not share the same eigenstates, the above terms will not cancel and the non-commutativity relation $Q_{n} Q_{m} - Q_{m} Q_{n} \neq 0$ holds. As $Q_{n} Q_{m} \neq Q_{m} Q_{n}$, the distribution in  \eqref{p_nm} that results from measuring $Q_{n} Q_{m}$ on $|\psi_{0}\rangle$ will differ from the distribution that results from considering the measurement order $Q_{m} Q_{n}$. Hence, the distribution over measurement results depends on the order of the observable measurements. Note that for $N$ observable measurements, there exist $2^N$ combinations of binary answers that the quantum system could reveal and $N!$ order permutations that could produce different binary distributions. 

Following \cite{guo2016noncommutativity}, for $N$ observables in the subset $\Gamma = \{Q_{n}: 1 \leq n \leq N\}$, the amount of non-commutativity of the total set can be measured by 

\begin{gather}
    \zeta(\Gamma):= \sum_{n < m} \lVert [Q_{n}, Q_{m}] \rVert_{\text{Tr}}, 
\end{gather}

where $\lVert.\rVert_{\text{Tr}}$ is the trace norm such that $\lVert Q \rVert_{\text{Tr}} = \text{Tr}\sqrt{Q^{\dagger}Q}$. As the models used in this study remain small in the number of observables, we are able to obtain an exact set of updated observables at each training step and compute this metric directly. Of course, this approach does not scale, and still, it is convenient enough for a small-scale investigation like ours.

\paragraph{Quantum Generative Model}\label{qcircuit}

\begin{figure*}
\includegraphics[width=\linewidth]{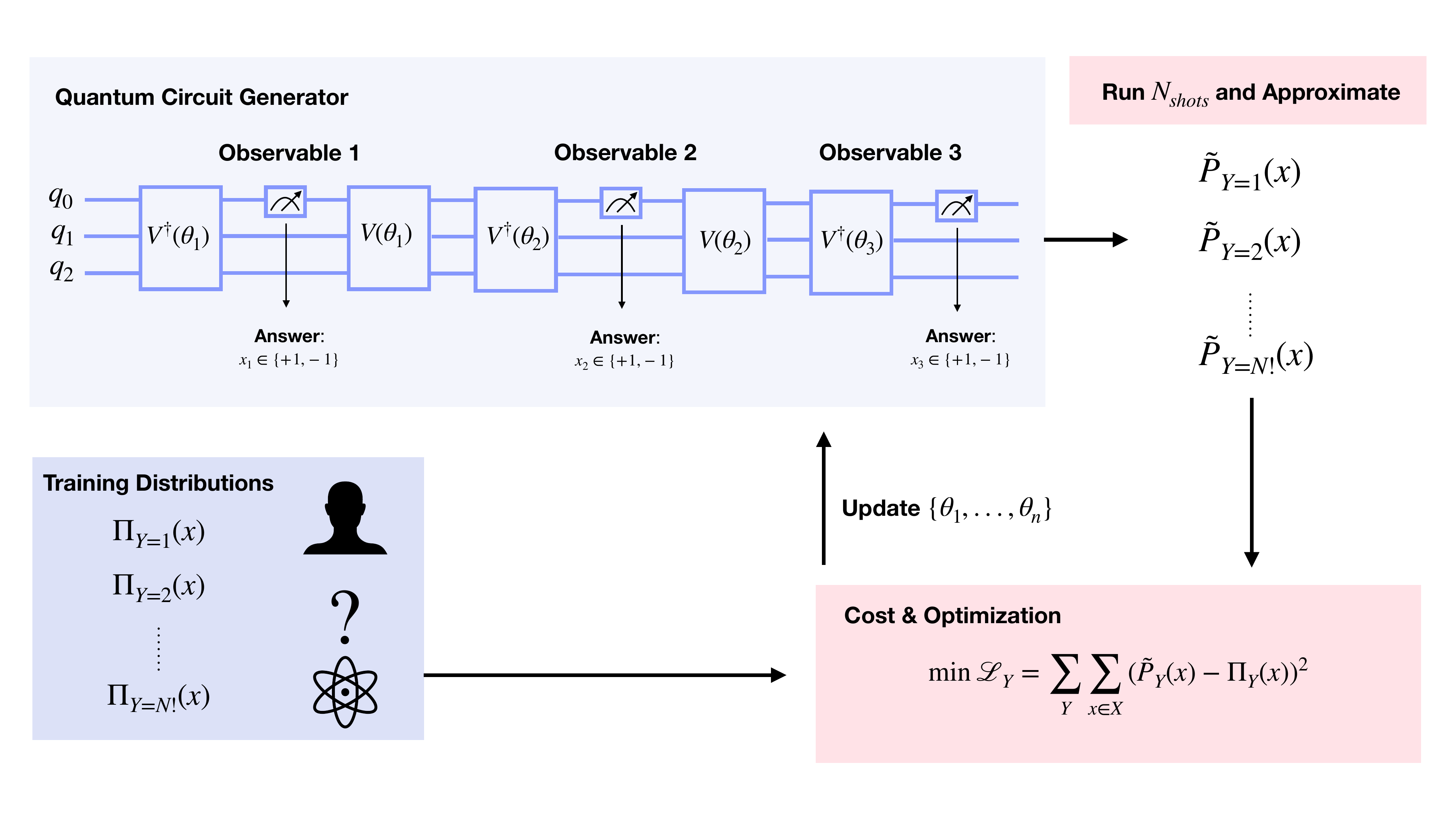}
\caption{\textbf{An example of the multi-task quantum generative architecture for training on binary distributions with three observables.} We provide a visual overview of the quantum circuit generator, containing subsequent measurement observables of the form $V(\theta_{n})D_{n}V(\theta_{n})^\dagger$. Measurements taken in the Pauli-Z basis allow one to obtain a binary eigenvalue (answer) for each trainable observable. The generator is run for a number of $N_{shots}$ to approximate the desired number of model output distributions $\tilde{P}_{Y}(x)$ over binary answer variables, with respect to the order sequence of the observables. Swapping the order sequence simply corresponds to physically swapping the parameters $\theta_{n}$ in the generator, which are indexed by $Y$. The model is trained to minimize the multi-task loss function $\mathcal{L}_{Y}$ that computes the LSM loss on the training distribution $\Pi_Y(x)$ and the empirical model distribution $\tilde{P}_Y(x)$ for the desired number of tasks to use for training. }
\label{fig:model}
\end{figure*}

Borrowing intuition from Section \ref{ind_bias}, we put forth an unsupervised quantum generative model that can learn distributions over binary strings $x \in \{-1, 1\}^{N}$ from measuring a sequence of $N$ quantum observables. We define an unsupervised generative model as one in which given a set of i.i.d. unlabeled datapoints from an unknown target distribution $P(x)$, the goal of the model is to learn the underlying features in the training samples such that the true $P(x)$ can be best approximated, and valuable data can be efficiently generated \cite{gili2023generalization, Bengio-book-2009}. Parameterized quantum circuits can be understood as interesting candidates for natural, universal approximators of probability distributions with efficient sampling capabilities, and have been investigated as generators for various ML tasks \cite{benedetti2019parameterized, Benedetti2019, rudolph2022synergistic, gili_qcbm, gili_qnbm, rudolphhandwritten}. 

We provide a visualization of our entire algorithm in Figure \ref{fig:model}. The quantum circuit requires the number of qubits $n_{\text{qubits}}$ to grow linearly with the number of observables $N$ in order to have full expressivity over all possible binary answers. Starting in an initial state $|\psi_{0}\rangle = |00 \dots 0_{N}\rangle$, we take measurements of subsequent trainable observables, discussed in Section \ref{ind_bias}, each taking the parameterized form $Q_{n} = V(\theta_{n})D_{n}V(\theta_{n})^\dagger$. In the circuit implementation, each unitary transformation $V(\theta_{n})$ is composed of arbitrary parameterized quantum gate sequences that play a role in the expressivity of the model. Specifically, we utilize fully expressive single-qubit rotation gates consisting of Pauli-$X$ and Pauli-$Z$ rotations, $R_{X}(\theta)R_{Z}(\theta)R_{X}(\theta)$ respectively on each of the qubits, with $R_{l}(\theta) = \text{exp}(\frac{-i \theta \sigma_{l}}{2})$. After each sequence of single-qubit operations, we utilize multi-qubit entangling gates, specifically $XX$ couplers in-between nearest-neighbor qubits, known as a line-topology \cite{Zhu2018}. The gate sequence for each observable $Q_n$   does not change when observables are permuted, and only depends on the parameters $\theta_{n}$. Thus, in our ansatz, the gate count and number of parameterized operations grows linearly with the number of qubits $n_{\text{qubits}}$ and the number of observables $N$. The total number of model parameters in the ansatz is given by $N(4n_{\text{qubits}}-1)$. Note that understanding the optimal gate sequences for $N$ observables is an open question for future work, especially in the regime that extends beyond binary data. 

After implementing $V(\theta_{n})$, we take a single (one-shot) measurement of the first qubit in the Pauli-Z basis to obtain the corresponding eigenvalue (binary answer) $x_{i} \in \{1,-1\}$. While more than one method can be utilized to simulate these sequential measurements, we implement a the circuit using a mixed state phase damping method that requires an additional qubit in our specific circuit implementation (code provided in \ref{code_implementation}). The shot samples from our quantum generator are then taken from the distribution
$P_{Y}(x)$, where $Y \in \{1, 2, ..., N!\}$ refers to the order of the observables (or, equivalently, the order of the ansatz parameters $\theta_n$) in the circuit. Depending on this order, the circuit gives rise to one of the $N!$ probability distributions and can therefore be used to learn different tasks. 

The model is trained on a task or order by minimizing the loss function $\mathcal{L}_Y$ between the empirical distribution $\tilde{P}_{{Y}}$ generated from the output samples and an empirical distribution over the training data $\Pi_{Y}$. Typical loss functions include distance measures such as the negative log likelihood (NLL) and the Maximum Mean Discrepancy (MMD) loss \cite{liu2018differentiable}. In this work, we utilize a simple Least Mean Squares (LMS) loss at each iteration defined as: 
\begin{equation}\label{lsm}
\mathcal{L}_Y= \sum_{x\in X} (\tilde{P}_{Y}(x) - \Pi_{Y}(x))^2,  
\end{equation}

We highlight that the question of how to best train generative quantum models is still open. Computing loss functions that require complete access to the model output probability distribution, like the NLL and LMS above, have exponential scaling relative to the dimension of the data samples and require good parameter initializations. As such, these loss functions are only useful in simple research settings like ours, where the data dimensions remain small. In more practical settings, the current best practice is to utilize loss functions based on f-divergences \cite{f_diverge} and maximum mean discrepancies \cite{rudolph2023trainability}.

Once computed, the loss is utilized to approximate the gradient in an Adam optimizer \cite{zhang2018improved}, which updates the parameter values and feeds them back into the circuit for the next iteration, until convergence is reached. 

\paragraph{Multi-Task Training Scheme}\label{multi-task}

Given a set of related learning tasks $\mathcal{T}$, multi-task learning aims to enhance a model on all tasks by using the knowledge contained in only a few of them \cite{multi-task_review, multi_task_generative}. The model usually consists of a base model and an adaptation to each task, such as a neural network whose weights are shared between all tasks for all but the final layer. Multi-task learning can be put in the context of generative modeling in the following way: a task $T_{i} \in \mathcal{T}$ consists of learning to sample from a probability distribution $P_{i}(x)$ given a finite set of training data from that distribution. The goal is to train on data from several tasks simultaneously with a loss function $\mathcal{L} = \sum_{i}\mathcal{L}_{i}(T_{i})$ and generate datapoints both from the array of tasks. This multi-task approach is expected to lead to a learning advantage over training separate architectures - either in terms of run-time, sample complexity, or generalization performance. 

In our learning setting, we can define each task $T_{i}$ as the quantum generator learning a distribution over binary answers with a fixed order of measurement observables. While the observables themselves are used for all tasks, their order -- in our case, the parameters of the ansatz -- gets adapted to the question order of a specific task. As such, our loss function for training becomes: 

\begin{equation}\label{nll}
\mathcal{L}_Y = \sum_{Y} \sum_{x\in X} (\tilde{P}_Y(x) - \Pi_{Y}(x))^2
\end{equation}

Similarly to training on a single distribution, one can use a gradient descent algorithm to minimize the loss function. Not all orders are required for training, and excluding orders from the training set to use in a test set is necessary to assess generalization performance to new tasks from $\mathcal{T}$. As the number of tasks required to compute the loss and achieve quality generalization performance for a scaled number of observables remains uncertain, the computational efficiency of the loss function is unclear.  

In addition, we want to highlight the distinction here between the model generalizing to \textit{tasks} (out-of-task generalization) vs. the model generalizing to \textit{samples} (out-of-distribution generalization) from each task. In classical multi-task learning, typically very few samples from all tasks are utilized with the goal of obtaining good out-of-distribution generalization (see for example, \cite{multi-task-small-samples}). Here, we are interested in out-of-task generalization as an evaluation metric to determine whether the bias of non-commutativity in the circuit model helps to learn the general structure of the tasks, i.e. the order effect underlying the distributions. We are using this model and scheme to obtain a more fundamental understanding as to how quantum mechanics plays a role in a learning setting and are not trying to claim computational advantage with our quantum model or training scheme. 

To summarize our quantum learning setting, we want to emphasize \textit{two main features} that result from the non-commutative nature of our quantum model and training scheme that intuitively argues its effectiveness for learning order effects in data: 

\begin{enumerate}
\item \textbf{A multi-task design}: The model can represent and train on multiple probability distributions at once - specifically those conditioned on the order of observables. Thus, observable order becomes an inherent variable of the model.

\item \textbf{A design that matches the data-generating process}: The physical mechanics of the model matches the physical process of the data generation. More specifically, obtaining information from a human in a survey corresponds to observing a property of a quantum system. 
\end{enumerate}

\paragraph{Psychology-Inspired Datasets}\label{data_set_artificial}

For our artificial datasets, we take inspiration from order effect studies that relate to human cognition and decision making \cite{measure_new_ordereffects, sequence_effects, wellbeing_order_effects}. In this context, a classical order effect can be defined as the change in distribution over human answers based on the order of questions asked. These order effects are prevalent in survey data across most disciplines and within most countries \cite{order_cultures, items_grids, wellbeing_order_effects}. In the psychology literature, we find many reasons for such order effects to occur. Typically, we see order effects in one of the following circumstances: (1) humans want to appear consistent with their answers such that their responses to previous questions will impact future answers; (2) humans don't want to be redundant so they try to provide new information in sequential answers; (3) humans use direct information or intuition from the previous question in their next answer (known as ``priming'') \cite{sequence_effects}. Many studies have shown these effects - ranging from asking questions regarding U.S. President likability, racism prevalence, abortion stances, and foreign affair policies \cite{item_order, ordereffects_germanystudy, response_order_effects}. 

In the third circumstance, order effects appear to occur most when two questions are asked sequentially on the same topic or on very similar topics. This is where we could see the ``judgement effect'', where questions contain the same object and previous judgements may serve as standards for later comparisons \cite{items_grids}. More specifically, many of these datasets follow the rule of ``evenhandedness'', which aims to explain the famous Al Gore/Bill Clinton example \cite{measure_new_ordereffects}. In this example, a population of people are asked \textit{Do you generally think Bill Clinton/Al Gore is honest and trustworthy?}. It is observed that respondents are more inclined to change their binary answer to the second person depending on the first person asked, such that the outcomes are ``more fair''. Thus, if the first question is about Clinton, people are more likely to lower their opinion of Gore to make things more ``even'' because Clinton generally has an overall lower score. If the first question is about Gore, people are more likely to raise their opinion of Clinton. 

We take inspiration from these insights in psychology to generate two artificial datasets, which we will refer to as \textit{D1} and \textit{D2} in our numerical results. The first is a dataset containing order effects dependent on general likeability scores of individuals. Given $N$ questions of \textit{Do you generally think this $i^{th}$ person is honest and trustworthy?}, we assign randomly generated likeability scores $S_{n} \in \{0, 1\}$, to each person in question, where general public likability increases as $S_{n} \rightarrow 1$. Thus, when there is generally a good opinion of the person in question, the probability of someone saying \textit{yes} to the question is high. We use a simple Markov model to produce $N!$ binary training datasets over various orders for $N$ people in question, providing the probabilities that a population of people answered $\{\text{No}, \text{Yes}\}$ in the sequence of questions. 
The model uses the likeability scores as hidden variable information, adjusting the outcome probabilities, by updating subsequent scores $S_{n+1}$ dependent on the current score $S_{n}$ such that it is either raised or lowered by the average difference between the two scores - thus making outcomes ``more even''. The relation for which an updated score $S'_{n+1}$ is computed is given by: 

\begin{equation}
    S'_{n+1} = S_{n+1} \pm \frac{S_{n+1} - S_{n}}{2}
\end{equation}

For our second dataset, we generate a sequence of general to specific ranking questions with a corresponding random list of probabilities for public answers. If the order is changed such that a more specific question precedes a more general question, the answer to the more general question will obtain a pre-defined $x\%$ increase in the probability that the answer is \textit{yes}. This dataset is based on authentic data regarding school bullying, where we see $45\%$ increases of students answering yes to being bullied if they are asked about a specific type of bullying first (e.g. cyber bullying) \cite{citation-key}.

In order to compute the strength of the order effect that exists within each dataset, we explicitly compute the LMS similarity metric (Equation \ref{lsm}) over all possible data combinations that result from changing the order (note again that this is only possible due to the small scale of our experiments). For the scope of this work, it is useful to define an Order Effect (OE) Strength  that allows us to quantify the amount of order effect that exists within our dataset for our model to learn.

\section{SUMMARY OF EXPERIMENTS} 

Having provided the background motivation for this investigation and the details of our model and training scheme, we now demonstrate our progress towards answering the following question: \textit{Does the non-commutative nature of quantum measurements make quantum models constructed from sequential measurements well-suited to learn datasets generated by order effects?} 

In our experiments we run a total of $15$ independent training runs for each experiment, each with different random seeds for training and data generation. The following plots report the average training loss with standard deviation error for $150$ epochs. We use an Adam gradient optimizer with $step size = 0.1$ \cite{Kingma2014}. Each circuit is implemented on a statevector simulator, outputting the model probability distribution directly. The experiments were implemented in the PennyLane software framework \cite{bergholm2022pennylane}. The experiments are designed to answer two sub-questions: 

\paragraph{\textbf{\textit{1. As we increase the strength of the order effect in the dataset, does the model learn a stronger degree of non-commutativity?}}}\label{non_commute_results}

To address this question, we trained a $N=2$ observable model on datasets D1 and D2 for increasing values of OE strengths, including datasets with no order effect present. For D1, the Order Effect (OE) strengths are controlled with likeability scores that are linearly increasing by $0.1$ for the second question, where the likeability score for the first question remains fixed at $0.1$ (for example, $[0.1, 0.2]$, $[0.1,0.3]$, $[0.1,0.4]$...). The larger the separation between scores, the larger the order effect. Note that for the likeability scores $[0.1, 0.1]$, there is no order effect. For D2, we linearly increase the variable $x\%$ in increments of $10\%$ from $\{0\%, ... 90\%\}$, where $0\%$ indicates an absence in order effect. These corresponding OE strengths for both datasets are computed from the resulting probability distributions as described in Section \ref{data_set_artificial}. 

\begin{figure}[h!]
\begin{subfigure}{.5\textwidth}
  \centering
  \captionsetup{justification=centering,margin=1cm}
  \includegraphics[width=0.95\linewidth]{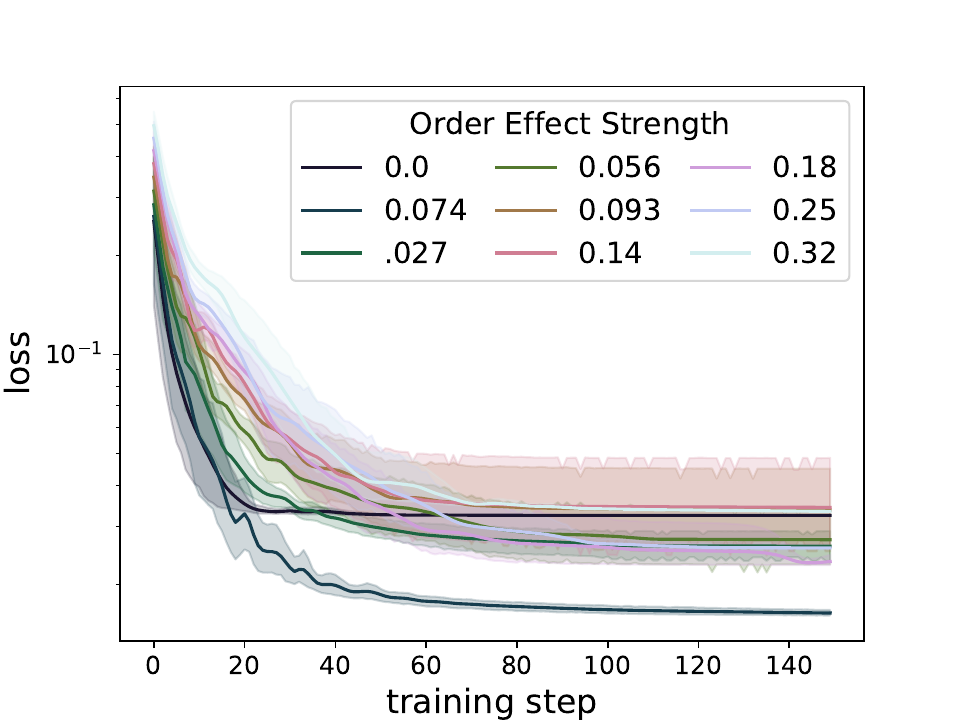}
  \caption{D1: LMS Training loss.}
  \label{fig:loss_D1}
\end{subfigure}
\begin{subfigure}{0.5\textwidth}
  \centering
  \captionsetup{justification=centering,margin=1cm}
  \includegraphics[width=0.95\linewidth]{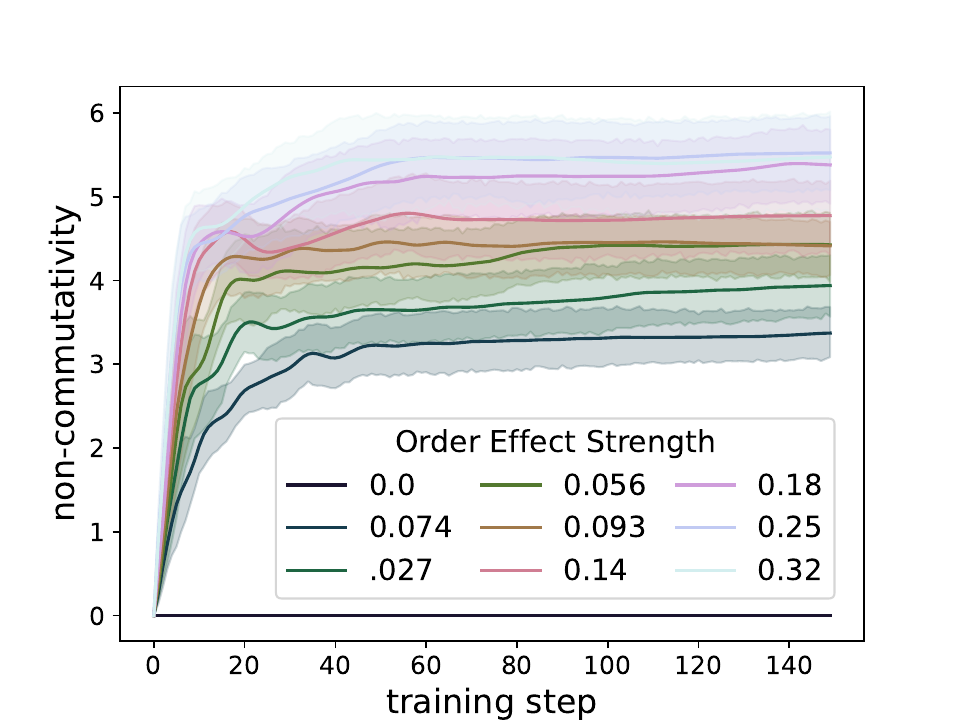}
  \caption{D1: Non-commutativity score.}
  \label{fig:noncommute_D1}
\end{subfigure}
\begin{subfigure}{.5\textwidth}
  \centering
\captionsetup{justification=centering,margin=1cm}
  \includegraphics[width=0.95\linewidth]{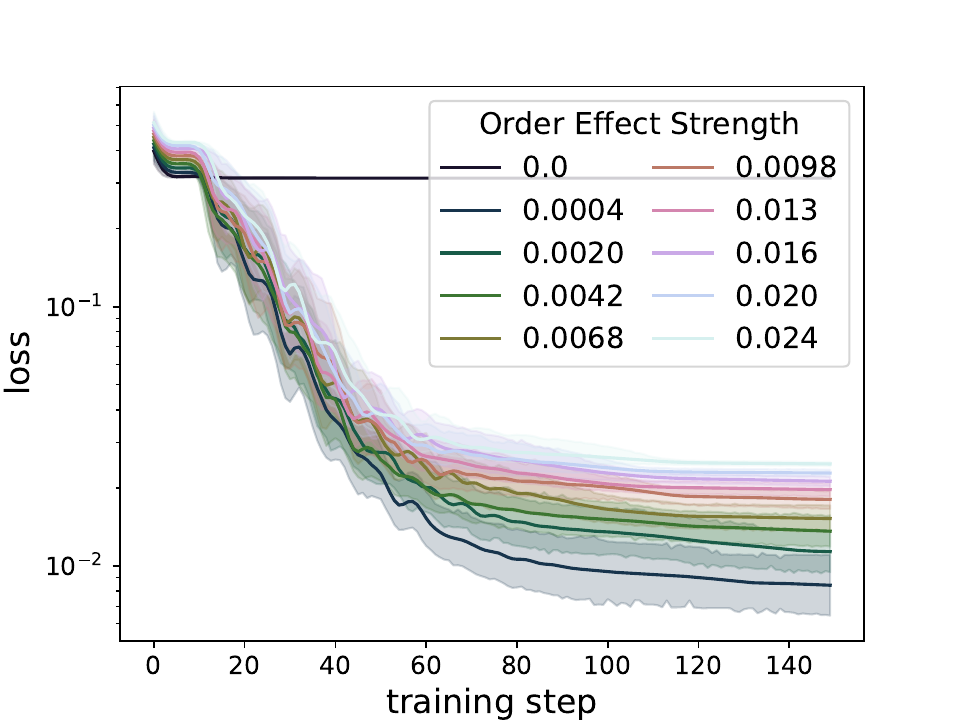}
  \caption{D2: LMS Training loss.}
  \label{fig:loss_D1}
\end{subfigure}
\begin{subfigure}{0.5\textwidth}
  \centering
  \captionsetup{justification=centering,margin=1cm}
  \includegraphics[width=0.95\linewidth]{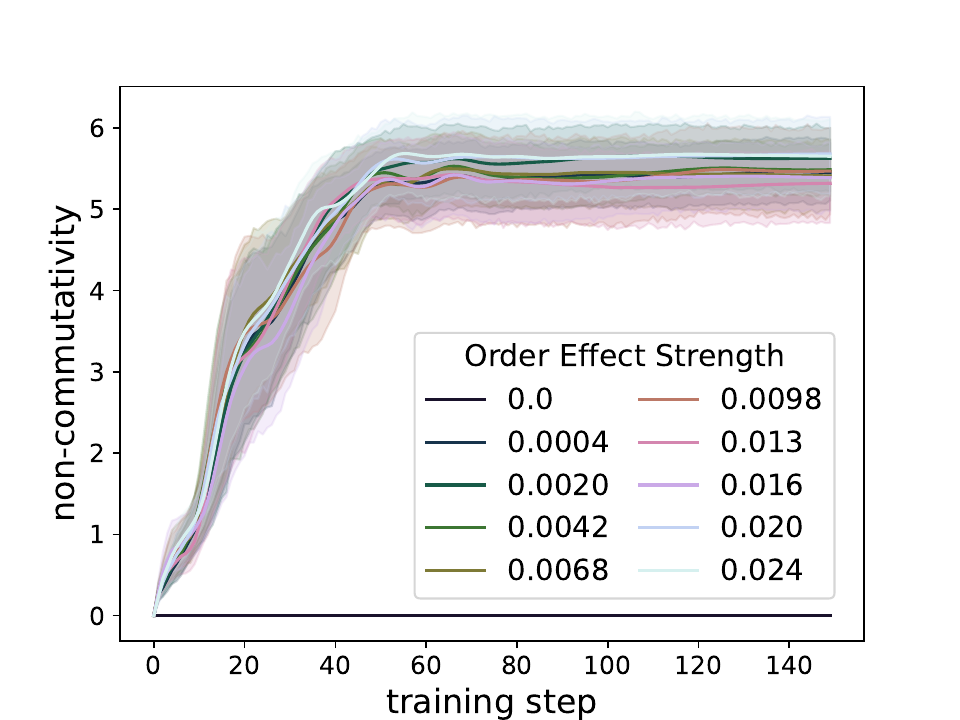}
  \caption{D2: Non-commutativity score.}
  \label{fig:noncommute_D1}
\end{subfigure}

\caption{\textbf{LMS loss and amount of non-commutativity present in the $N=2$ observable model throughout training for datasets D1 and D2 with various Order Effect strengths.} Each figure contains resulted obtained over $15$ independent trials with random seeds for train and test data, as well as randomness within the optimization. In (a) and (b), we see very clearly that as one increases the strength of the order effect in the dataset, the model learns observables with a higher degree of non-commutativity. There is not evidence of this in (c) and (d) for D2, most likely because the OE strengths are already a lot smaller than in D1. As the model is \textit{trained} to learn the order effects, we see the observables learn a larger degree of non-commutativity for both D1 and D2 in all cases, except for when there is no order effect. 
}
\label{fig:non_commute_data}
\end{figure}

We intentionally started training with two identical, and hence commuting observables, to provide a ground-zero starting point. There is no out-of-task generalization present in these experiments, as we train on samples from different orders or tasks and test on others. The training results with respect to the LMS loss and the amount of non-commutativity learned in the dataset are shown in Figure \ref{fig:non_commute_data}. For both experiments, the model learns an increasing amount of non-commutativity throughout training, except in the absence of an order effect in the data. This suggests that the model uses non-commutativity as a resource to solve the tasks.  

As shown in Figure \ref{fig:non_commute_data}, in D1 it is very clear that the model learns a higher degree of non-commutativity as we increase the OE strength in the dataset. This is not evident in D2. We hypothesize that in D2, the OE strengths are so small ($\leq 0.03$) that the dataset does not have enough bias for the model to learn. We question whether OE strengths of that size are negligible for the model, and perhaps order effects in the dataset are bounded by a certain strength for the model to be able to learn them. We discuss this further in the outlook as a direction for future investigation. 

For both datasets, a non-zero low strength order effect in the dataset is easiest to train on, but does not correlate to as high of a non-commutativity score. In fact, there is no non-commutativity learned in the model when no order effect is present. In addition, the highest strength order effect is one of the most challenging for the model to train on, and yet, achieves the highest amount of non-commutativity. This is not necessarily surprising since distributions with greater dissimilarity pose a larger challenge for the model to find a good set of parameters. It is interesting that the model still finds a good solution with a relatively low LMS loss, and that this directly relates to the model learning a higher degree of non-commutativity. The findings support an affirmative answer to the fist question, namely that the model \emph{does use} non-commutativity to learn the order effect.

\paragraph{\textbf{\textit{2. Can the model generalize to tasks with unseen question orders?}}}\label{generalization_results}

To gain some first insights into this question, we train models with $N = \{3, 4, 5\}$ observables\footnote{Note that the circuit corresponding to a 5-observable experiment uses 10 qubits on a mixed state simulator. Larger simulations are possible with today's software but still very time-consuming, and were out of the scope of this investigation.} on various portions of tasks (orders) with dataset D1. This time we use randomly selected likeability scores to create the tasks that comprise the test and training sets. When using a limited amount of tasks for training, we keep the number of test orders fixed at $10$, except in the case where $N=3$ and there are only 6 possible tasks.  These test tasks are randomly selected over the 15 trials, along with the training tasks. We also introduce randomness into the optimization over each trial.  If the model's test loss on unseen tasks decreases with its training loss on training tasks, we count this as evidence that the model is out-of-task generalizing. Below, we incorporate a metric called the \textit{generalization difference}, which is the difference between the average LMS test loss at the beginning and end of training. A greater decrease in the test error, over the averaged set of tasks, indicates that the model is better at learning from the data. 

\begin{figure}[h]
\begin{subfigure}{.5\textwidth}
  \centering
  \captionsetup{justification=centering,margin=1cm}
  \includegraphics[width=1\linewidth]{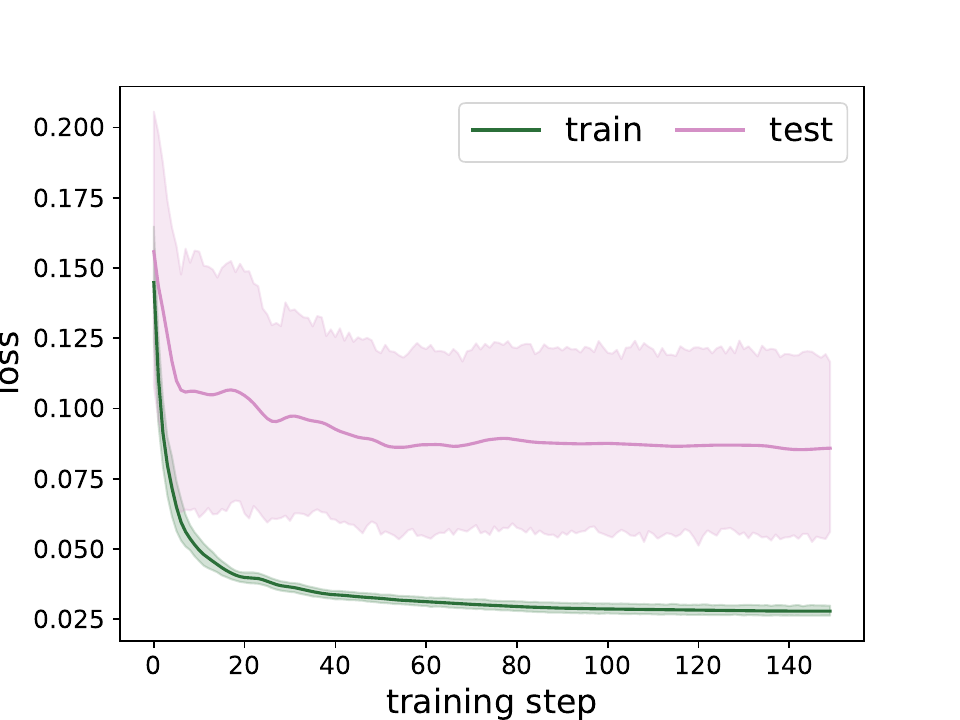}
  \caption{Train and test loss for 3 observables trained on 5 orders.}
  \label{fig:gen_3}
\end{subfigure}
\begin{subfigure}{0.5\textwidth}
  \centering
  \captionsetup{justification=centering,margin=1cm}
  \includegraphics[width=1\linewidth]{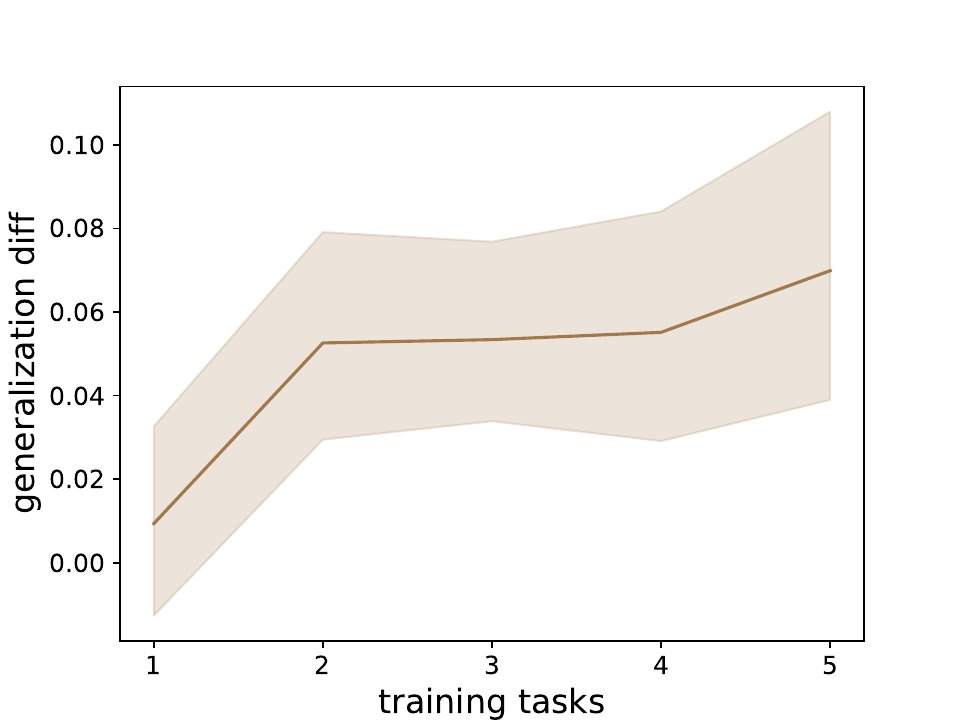}
  \caption{Generalization difference when trained on increasing numbers of tasks.}
  \label{fig:gen_dif_3}
\end{subfigure}
\caption{\textbf{Numerical results for $N=3$ observables on D1.} On the left, show an example of the decrease in train/test loss, such that one can visualize the generalization difference when trained on a specific number of tasks. On the right, we show the generalization difference with respect to training on an increased number of tasks. Each plot shows the average of $15$ independent trials from randomized train/test data and optimizer exploration.  We see that the model's test loss is initially congruent with the training loss, indicating that the model is able to generalize from seen to unseen tasks even at small scale. However, the test loss does not reach the same performance as the training loss. As we train on more orders, the model is able to generalize more - which is theoretically expected behavior. 
}
\label{fig:generalization_3_observables}
\end{figure}

In Figure \ref{fig:generalization_3_observables}, we show the results for $N=3$ observables when trained on various orders. Note that this is a very small-scale example with only $6$ possible tasks in total, which allows us to run experiments that train on one task to all tasks \textit{but} one. For all experiments we see that the model's test loss on unseen tasks decreases alongside the training loss. As expected, we see that the model's generalization difference increases as we train on more orders: the model has access to more tasks, and can therefore generalize more effectively. When trained on $5$ tasks and generalizing to a single task, the model's test loss on unseen tasks is closest to the training loss. However, it is also apparent that not all information about the unseen tasks is learned, since training and test loss differ significantly.

\begin{figure}[h]
\begin{subfigure}{.5\textwidth}
  \centering
  \captionsetup{justification=centering,margin=1cm}
  \includegraphics[width=1\linewidth]{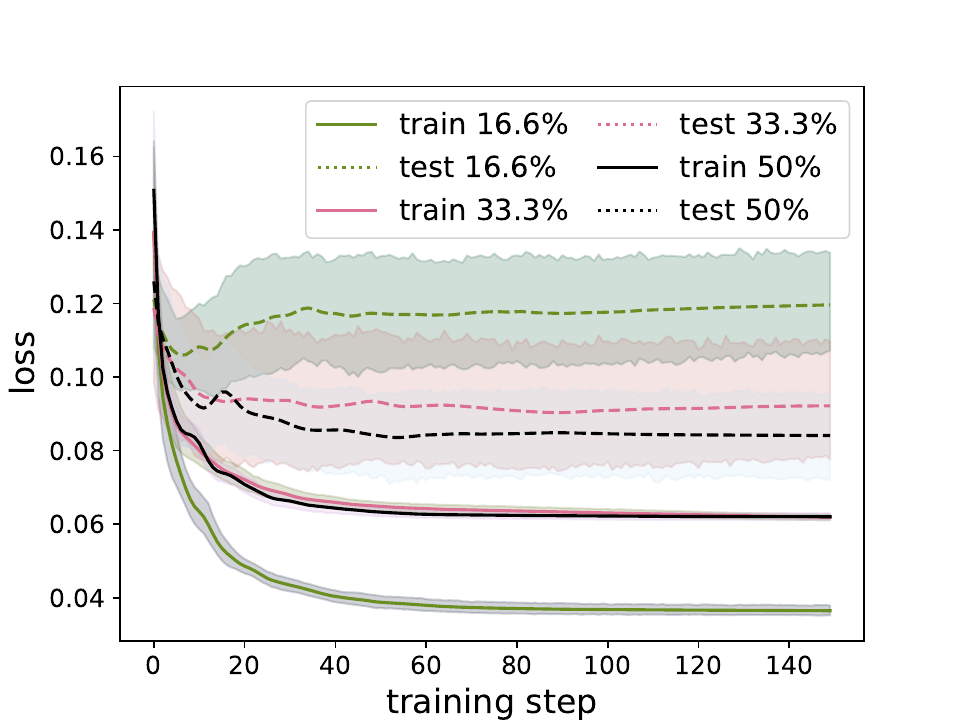}
  \caption{Train and test loss for 4 observables trained on task percentages.}
  \label{fig:gen_4}
\end{subfigure}
\begin{subfigure}{0.5\textwidth}
  \centering
  \captionsetup{justification=centering,margin=1cm}
  \includegraphics[width=1\linewidth]{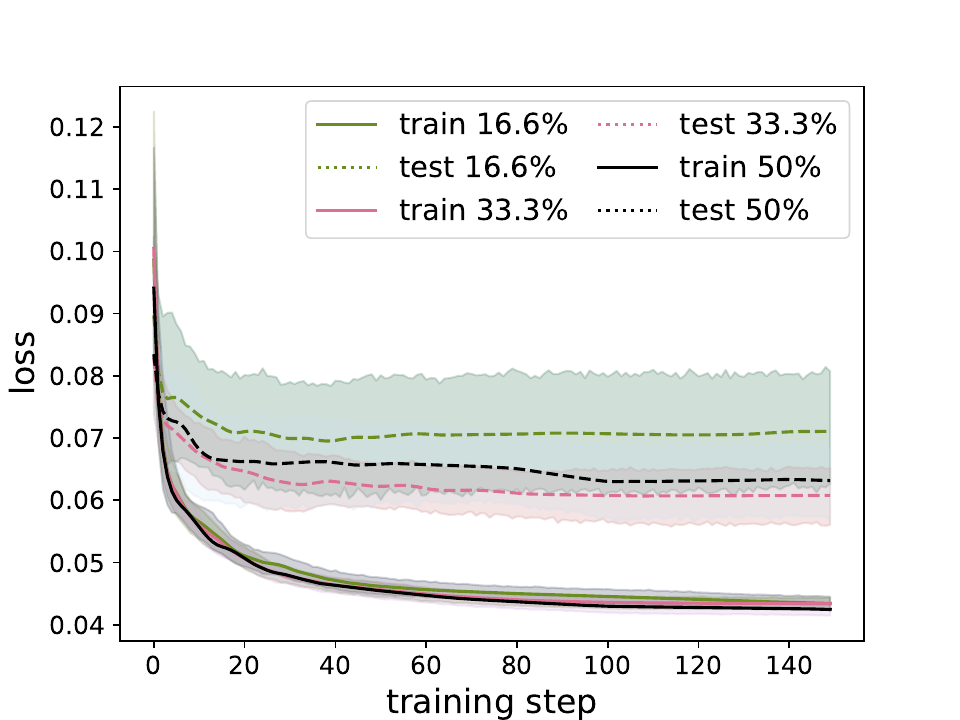}
  \caption{Train and test loss for 5 observables trained on task percentages.}
  \label{fig:gen_5}
\end{subfigure}
\caption{\textbf{Numerical results for $N=4$ and  $N=5$ observables on D1.} We show average results over $15$ independent random trials with respect to the train/test data and the optimization process. In both cases, the model is able to generalize to unseen orders. It is clear for $N=4$ observables that as we increase the number of tasks shown to the model, the model's test error gets lower. For $N=5$ observables this is slightly less clear, and perhaps because the problem size is larger and hence more difficult for the model to navigate a larger parameter space. However, we also see that in the case of $N=5$, the model is able to obtain better test errors than in the case of $N=4$. 
}
\label{fig:generalization_4_5)observables}
\end{figure}

When scaling our experiments to $N=4$ and $N=5$ observables, we do not see as clear of a trend in generalization difference as we increase the number of tasks during training. As there are more potential orders to generalize to ($4$ observables: $24$ orders, $5$ observables: $120$ orders), we randomly selected $16.6\%$, $33.3\%$ and $50\%$ of all tasks. The corresponding training and test losses are displayed in Figure \ref{fig:generalization_4_5)observables}.

In both cases, the model learns to generalize to some extent to unseen orders, which indicates a weak \textit{yes} to our research question. However, there is nuance in discussing the quality of the generalization performance and the amount of data required to achieve such quality. It is clear for $N=4$ observables that as we increase the number of tasks shown to the model, the model's test error gets lower and closer to the training error, which means that generalization improves.  Still, we recognize that the test error does not reach the same low loss as the training error, indicating that the out-of-task generalization performance result is weak. 

For $N=5$ observables the trend is less clear, and perhaps because the problem size is larger and hence more difficult for the model to navigate a larger parameter space. As seen in other practical ML demonstrations \cite{gili2023generalization}, it is possible that we are hitting a \textit{Goldilocks region}, where increasing the amount of data shown to the model actually makes it more difficult to generalize. We also see that in the case of $N=5$, the model is able to obtain better test errors than in the case of $N=4$, an observation that would be interesting to study as the number of observables scales further. 

Note that the experiments do not allow conclusions about the model's out-of-distribution generalization capabilities within a task, nor about whether or not the non-commutative structure reduces the amount of orders that we need to train on when scaling the problem size (which would be an interesting investigation for future work). What we can conclude from the experiments is that non-commutativity is used as a resource, and that the model design allows it to use information gained in the training on some tasks to produce better samples from others, a sign that the inductive bias of the model may be useful for the learning task.

\section{DISCUSSION AND OUTLOOK}\label{s:outlook}

As the field of QML is quickly expanding with a unified intention of discovering applications for quantum computers, it is not surprising that there now exists a plethora of approaches to reach this outcome. The most popular research approach of the last few years borrows language and tools from computational complexity theory to look for quantum learning settings with a built-in computational speed-up \cite{hinsche2021learnability, hinsche2022, 2020Du, Liu_2021_speed, Yoo_2014, schreiber2022classical, cai2022sample}. This \textit{algorithmic quantum advantage} approach focuses on demonstrating that a programmable quantum device can solve a machine learning problem that no classical computer can solve in any feasible amount of time. A complimentary second approach has recently been introduced to look for quantum learning settings that contain industry relevance, where asymptotic speed-up may not be the primary measure of value, but instead the model's ability to generalize \cite{gili2023generalization, hibatallah2023framework, gili_qcbm, moussa2023application}. This \textit{practical quantum advantage} approach focuses on demonstrating that a quantum algorithm or heuristic can outperform a state-of-the-art classical method for a real-world ML problem \cite{herrmann2023quantum}. Emphasizing quantum learning settings with trainability guarantees is an adjacent third approach in which one tries to obtain efficient training methods for more general models \cite{abbas2023quantum, rudolph2022synergistic, rudolph2023trainability}. We note that all of these approaches deliver important insights to the field of QML. 

In this work, we exercise an alternative approach that focuses on finding quantum learning settings where the mathematical structure of quantum mechanics leads to an inductive bias in a quantum model that matches the structure it has to learn from a particular kind of dataset \cite{bowles2023contextuality, kübler2021inductive}. This approach aims to increase our overall \textit{understanding} of the potential natural alignments between quantum models and data structures that would lead to advantage. As for the specific scope of our work, we contribute a quantum learning setting that allows one to study an inductive bias that results from the non-commutative nature of quantum measurement. By introducing artificial datasets inspired by human psychology and a multi-task quantum generative model that contains a theoretical bias towards learning order effects, we investigate in our numerical experiments whether this bias helps the model learn datasets with increasing order effect strength and aids generalization performance. We find some initial evidence that the non-commutative bias of the model is indeed useful for learning order effects in data. We see this as a valuable insight for both the Quantum Cognition and QML community. Still, more evidence is necessary to gain clarity on how the resource of non-commutativity can be used in different settings. Possible questions are: \textit{How do our findings hold up when scaling the tasks?}\textit{Can we lower the out-of-task generalization error further?}\textit{Does the model's non-commutative bias remain effective when the order effect in the dataset is small?} \textit{Can this learning setting be adapted to isolate non-commutativity as the primary variable that leads to improved generalization performance?} \textit{Does the quantum model suit other types of datasets?} Most of all, we hope this work inspires other researchers to consider alternative ways in which we can  investigate the relationship between quantum mechanics and learning. 

\section{Acknowledgements}
K.G. would like to acknowledge all of the behind the scenes work that provided time for this research to be completed, as well as the support from her closest friends and mentors who continuously challenge her to think differently. 

\section{Code Implementation} \label{code_implementation}

The code for this work can be found in the public repository: \url{https://github.com/kaitlinmgili/noncommutativity-ordereffects.}

\printbibliography

\end{document}